\documentclass[preprint,showpacs,preprintnumbers,amsmath,amssymb,superscriptaddress,floatfix]{revtex4}

\usepackage[dvipdfmx]{graphicx,color}

\usepackage{dcolumn}
\usepackage{bm}
\usepackage{color}
\usepackage{hyperref}
\usepackage{array}

\begin{document}

\title{A significantly stable mode of the ultracold atomic wave packet in amplitude modulated parabolic optical lattices}
\noindent
\author{Tomotake Yamakoshi}
\affiliation{Institute for Laser Science, University of Electro-Communications, 1-5-1 Chofugaoka, Chofu-shi, Tokyo 182-8585, Japan}
\author{Farhan Saif}
\affiliation{Department of Physics and Electronics, Quaid-i-Azam University, Islamabad 45320, Pakistan}
\author{Shinichi Watanabe}
\affiliation{Department of Engineering Science, University of Electro-Communications, 1-5-1 Chofugaoka, Chofu-shi, Tokyo 182-8585, Japan}

\date{\today}

\begin{abstract}

We show that a conspicuous wave packet of ultracold noninteracting Bosonic atoms emerges in a 1-dimensional parabolic optical lattice as in the setup of the Aarhus experiment [P.~L.~Pedersen {\it et al.}, Phys. Rev. A {\bf 88}, 023620 (2013)], given the lattice height is harmonically modulated with a particular amplitude at a resonant frequency. 
We show that this wave packet, coined ``{\it 4bandPWP}'' here, executes stable time-wise periodic motion for infinitely long time.
We apply the Floquet theory to analyze the parameter dependence of {\it 4bandPWP} in detail.  
Our analysis shows that it consists mainly of two principal Floquet eigenstates of the periodically driven Hamiltonian. 
The informative Husimi representation yields temporal slices of the phase space of {\it 4bandPWP}, visually identifying moments where the inter-band transitions take place.
The provided data should aid the experiment in locating {\it 4bandPWP}.

\end{abstract}

\maketitle

\section{introduction}
\label{sect:Introduction}

Controlling ultracold atomic systems with time-wise periodic external fields has attracted attention particularly since the experimental realization of atomic Bose-Einstein condensation (BEC).  An enormous volume of ingenuous applications have been made possible in recent years (see Ref.~\cite{applications} for some examples and references therein).   Realizing tunable gauge fields\cite{gauge} like the Hofstadter Hamiltonian\cite{Hof} may be regarded as a striking example of wave packet manipulation\cite{wp-manipu}. 
The amplitude modulated optical lattice(OL) serves as a tool to generate an excited atomic wave packet from the ground state. And in combination with the weak parabolic trap, it serves to manipulate the so-generated wave packet.

Experiments by the Aarhus group\cite{Aarhus} and the Hamburg group\cite{Hamburg} were aimed partially toward this end, and their findings were theoretically accounted for in Ref's.~\cite{my-1,my-2}. 
When harmonic amplitude modulation is superimposed on the OL, the wave packet undergoes, in essence, inter-band transition to higher energy bands. 
And then in the course of propagating in time in the ``{\it static}'' OL plus the parabolic trap, it reveals complex features due to the Bragg reflection and the Landau-Zener transition across a band gap. 
Here, we may view that the free motion of the wave packet is governed by the energy-quasi-momentum dispersion while the force is exerted by the parabolic trap. 
The excited wave packet may travel to reach some spatial areas far from the trap center. 
De-excitation by a second application of the amplitude modulation at this location may bring the wave packet into localized eigenstates.
Indeed, the experiment at Aarhus\cite{Aarhus} employed amplitude modulation to selectively excite ultracold atoms to a specific band and succeeded in transferring them to spatially localized eigenstates.

It appears natural to extend the analysis of the wave packet from the case of {\it static} OL system to the case of {\it continuously driven} OL\cite{rec-track,atom-int}. 
In understanding characteristics of the wave packet dynamics under a continuous drive, we explored wave packets for various values of the modulation amplitude and  frequency.
In so doing, we encountered one conspicuously long-lived stable ``periodic'' wave packet.  It is important to realize that this is a particular Floquet state, the system being time-wise periodic. 
A schematic representation of the phenomenon is given in Fig.~\ref{fig:time-independent}, indicating the locations of resonant transition by arrows. 
These resonances are what marks the major difference from the static parabolic OL.  
Since interband transitions involving four bands occur, let us refer to this periodic wave packet as the 4 band periodic wave packet, or ``{\it 4bandPWP}'' in short, from here on. 
It turns out that a similar phenomenon was previously reported in Ref.~\cite{two-Flo} where a resonant transition occurs between a pair of initially degenerate eigenstates. 
Their degeneracy gets lifted by ``dynamical tunneling''. 
The situation is akin to the periodic transition of a wave packet from one well to the other in the familiar double well problem. 
Let us call it the ``{\it tunneling}'' wave packet for short. 

It is not difficult to imagine multitudes of periodic wave packets in the current system, identifiable as Floquet states, the two above being mere examples. 
However, not all the Floquet states are reachable in a particular experiment because the initial condition differs from experiment to experiment. 
Indeed, {\it 4bandPWP} and the {\it tunneling} wave packet belong to different initial conditions. 
To create the {\it tunneling} wave packet, an appropriate phase shift is first applied to the ground-band atoms to impart momentum to the wave packet, and cast it into the region corresponding to a stable island in classical phase space\cite{two-Flo}. 
 There are a couple of more differences.
In the case of  the {\it tunneling} wave packet, the frequency of amplitude modulation exceeds the lattice height so that the wave packet is sent to the quasi-continuum state, obviating the concept of energy bands. In the case of {\it 4bandPWP}, transition occurs between bands, {\it i.e.} the transition is very inter-band. The other difference is the absence of the parabolic potential in Ref.~\cite{two-Flo} so that the evolution of the system is merely due to tunneling under diffusion while  {\it 4bandPWP} evolves along the energy-quasi-momentum dispersion curve under the external force of the parabolic trap. 
This type of difference in experimental setup is an important aspect to bear in mind.

Let us also note that Ref.~\cite{saif} is one of the first papers that has presented the existence of various periodicities at various time scales subject to periodic drive and noted their importance to probing quantum chaos. 
The foundation concept was tested and confirmed experimentally~\cite{two-Flo}. 
For that matter, the present {\it 4bandPWP} is an important step to explain the dynamical characteristics of matter waves in the parabolic OL subject to periodic external amplitude modulation.

In this paper, we focus our attention on the physics of {\it 4bandPWP} as a nontrivial example of stable periodic wave packets that originate from the ground state of the static parabolic OL. 
It is conceivably beyond analytical methods to construct the time-dependent wave packet in a closed form.
Therefore we perform rigorous numerical simulation by solving the time dependent Schr\"odinger equation (TDSE) instead. 
In order to analyze the TDSE results, we exploit the Floquet theory within a limited subset of the Hilbert space as well as a semi-classical theory developed in Ref.\cite{Hamburg}.
This analysis allows us to estimate under what condition {\it 4bandPWP} manifests itself, and also to assess its stability.
Subsequently, we introduce a simple four-level model to approximately describe {\it 4bandPWP} and discuss its parameter dependence for the sake of concreteness. Here the four levels pertain to those eigenstates of the static OL system whose pair of linear combinations yield an approximate Floquet representation of {\it 4bandPWP}. The pair of states are analogous to that of the previously-mentioned {\it tunneling} wave packet, and are stable by design.

The paper is organized as follows.
Sect.~\ref{sect:system} outlines the system and its theoretical model. 
Sect.~\ref{sect:numerics} analyzes the numerical results of the long-time dynamics in detail.
The Husimi representation will be employed to gain insight into the correspondence with quantum dynamics.
To aid the experiment in locating {\it 4bandPWP}, we supply some numerical data.
Sect.~\ref{sect:conclusions} concludes the paper.

\section{Mathematical Definition of the System}
\label{sect:system}
We consider dynamics of an ideal Boson in the amplitude modulated parabolic optical lattice.
Some notations and techniques used in this paper are available in the experimental works of \cite{Aarhus,Hamburg}, and in the recent numerical studies of \cite{my-1,my-2}. 
We use recoil energy $E_r=\hbar^2 k_{r}^2/2 m$ as the unit of energy, recoil momentum $k_{r}=2\pi/\lambda$ as the unit of (quasi-)momentum, lattice constant $a=\lambda/2$ as the unit of length and rescaled time $t=E_r t'/\hbar$ as the unit of time. Here $\hbar$, $\lambda$ and $m$ correspond to the Planck constant, laser wave length of the optical lattice, and mass of the particle, respectively. 

The 1D version of the system is described by the time-dependent Hamiltonian
$$H=-\frac{\hbar^2}{2m}\frac{\partial^2}{\partial x'^2}+ V_0\sin^2 (k_{r}x') [1+\epsilon_0 \cos(\omega ' t')] + \frac{1}{2}m\omega_0^2 x'^2$$ where $V_0$ is the height of the optical lattice, $\omega^\prime$ is the frequency of the amplitude modulation, $\epsilon_0$ is the modulation strength and $\omega_0$  determines the curvature of the trap potential. Rescaling the Hamiltonian, we get
\begin{eqnarray}
H &=& -\frac{\partial^2}{\partial x^2}+s \sin^2 (x) [1+\epsilon_0 \cos(E_\omega t)]+ \nu x^2  \nonumber \\ 
  &=& H_0 +s \sin^2 (x)\epsilon_0 \cos(E_\omega t)
\label{eq:re-ham}
\end{eqnarray}
where $x$, $s$, $E_\omega$ and $\nu$ denote $x=k_{r}x'$, $s=V_0/E_r$, $E_\omega=\hbar \omega ' / E_r$, $\nu=m\omega_0^2/2E_r k_{r}^2$, respectively so that $H_0=-\frac{\partial^2}{\partial x^2}+ s \sin^2 (x)+ \nu x^2$ is the static part of the OL Hamiltonian. 
The parameter $s$ gives depth of the optical lattice in the units of recoil energy. 
In experiments on ultracold atomic systems, this parameter can be easily controlled; in the Hamburg experiment\cite{Hamburg}, its typical value varies in the range of $s=2-20$.
In what follows, we solve the TDSE $i \frac{d}{dt} \Psi(x,t) = H \Psi(x,t)$ by rigorous numerical method.
We use $s=16$, $\nu=1.63 \times 10^{-4}$(70Hz) and $\epsilon_0=0.165$ unless otherwise noted. Throughout the paper, the initial state is taken to be the ground state of $H_0$ and the amplitude modulation suddenly turns on at $t=0$ as in the Aarhus experiment.

\section{Discussions of Numerical Results}
\label{sect:numerics}
\subsection{Analysis based on Time-Independent eigenstates}
\begin{figure}[htbp]
 \begin{center}
 \includegraphics[width=7cm]{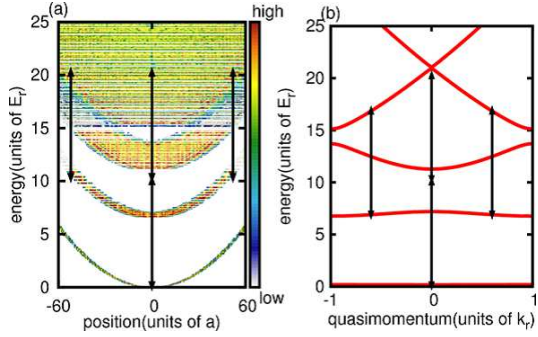}
 \end{center}
 \caption{(Color online) (a) Eigenstates of $H_0$ in position space in the energy range between 0 and 25$E_r$.
 The horizontal axis represents position coordinate $x$, and the vertical axis energy. Density becomes dense toward red and low  toward white.
 (b) The band structure deriving from $H_B$. We note that the ground energy of $H_0$ and $H_B$ are both set to 0.
 Double headed arrows show where bands couple resonantly with $E_\omega=$10.2. See Fig.~\ref{fig:husimi} below for the Husimi representation of this cycle.
 }
 \label{fig:time-independent}
\end{figure}

First, we grasp general features of the long-time dynamics by analyzing eigenstates of $H_0$.
Fig.~\ref{fig:time-independent}(a) shows the density of eigenstates in position space.
With the aid of the semi-classical theory\cite{Hamburg} and Bloch's theorem, main features of eigenstates can be explained.
The classical Hamiltonian under the single-band approximation is given by $H_{cl}=E^{n}_{q}+\nu x^2$ where $E^{n}_{q}$ corresponds to the band dispersion (Fig.~\ref{fig:time-independent}(b)) of Hamiltonian $H_{B}=-\frac{\partial^2}{\partial x^2}+s \sin^2 (x)$.
Here $q$ and $n$ are quasimomentum and band index of $H_B$, respectively.
Concerning the band indexes, we label the bands as the ground, first, second, ....
The classical Hamiltonian has two distinct regions in phase space separated by a separatrix\cite{Kolovsky}, namely, the Dipole mode and the Bloch mode. Analogy may be drawn to the oscillation mode and to the rotation mode of the classical pendulum, respectively with one notable difference that the roles of position and momentum appear {\it interchanged}.
The dipole mode occurs around the origin of the harmonic trap where $E^{n}_{q}$ is comparable to $\nu x^2$ so that the tunneling from one site to another is essential for characterizing the energy-band structure.
However, in the region of the Bloch mode, the harmonic potential dominates so that $E^{n}_{q}$ may be ignored and thus the tunneling is suppressed.
As a result, the eigenstates are spatially localized in this region.

\begin{figure}[htbp]
 \begin{center}
 \includegraphics[width=7cm]{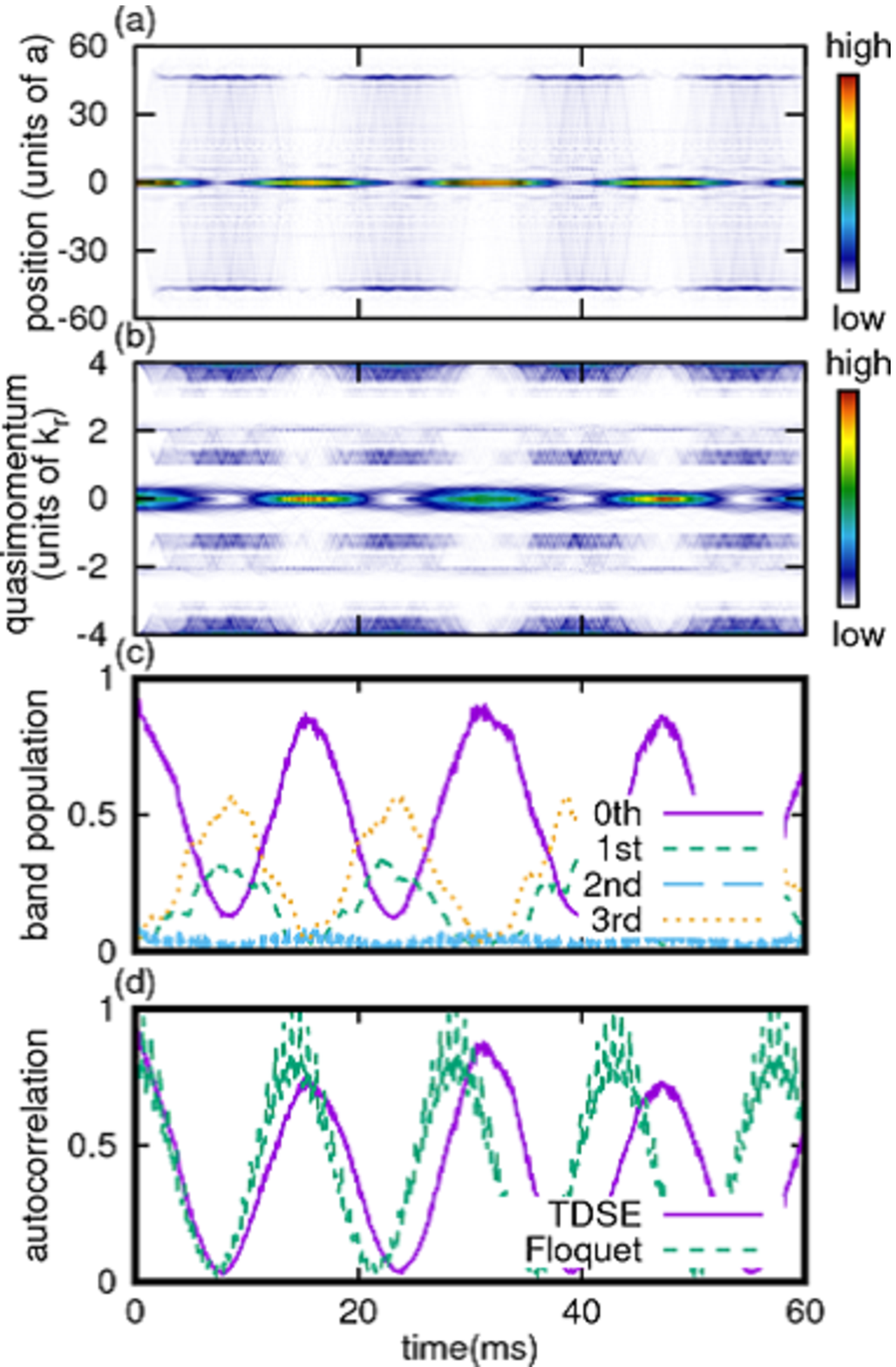}
 \end{center}
 \caption{(Color online)
 Density plots of ``{\it 4bandPWP}'' are shown in (a)~position and (b)~quasimomentum space.
 (c) shows the band population. 
 Purple solid line, green short-dashed line, light blue long-dashed line and orange dots correspond to ground(0th), 1st, 2nd and 3rd bands, respectively.
 (d) shows the autocorrelation function for TDSE (purple solid) and Floquet theory (green dashed).
 Floquet states discussed in subsection \ref{sect:floquet}.
 }
 \label{fig:time-evolution}
\end{figure}

The excitation process can be also analyzed by the time-independent approach as in Ref.~\cite{my-1}.
This preceding perturbative analysis of inter-band transitions indicates that the amplitude modulation only allows vertical transitions in the band dispersion of $H_B$.
A semi-analytic expression of ground state of $H_0$ in terms of quasimomentum is given by $Ae^{-q^2 \pi J^{1/2}/2 \nu^{1/2}}$ where $A$ and $J$ correspond to the normalization coefficient and the ground-band hopping parameter\cite{my-1}, respectively.
Our analysis throughout this paper thus concerns this ground state localized closely around $q=0$ as the initial condition.

Let us discuss how the preceding wave packet of our interest emerges. As shown in Fig.~\ref{fig:time-independent}(b), if we tune the double modulation frequency to the energy difference between ground and top of the 3rd-band ($E_\omega \simeq$10.2), the amplitude modulation produces two major couplings, namely, one between ground (dipole) and 3rd (dipole) via a two-photon process, and the other between 1st (Bloch) and 3rd (dipole) via a one-photon process. Here the term ``photon'' refers to an energy quantum of  $\hbar\omega^\prime$ emitted or absorbed due to the amplitude modulation rather than a real photonic transition.
In Fig.~\ref{fig:time-independent}(a), we show these major couplings also in position space.
Here it is important to recognize that the system could be treated as a closed system since no exact resonance exists between 3rd and any higher band.
Fig.~\ref{fig:time-evolution}(a) and (b) show density distributions of time-dependent wave function in position and quasimomentum space, respectively; the figures show periodic structure in both spaces.
Fig.~\ref{fig:time-evolution}(c) shows band population.
Populations in 1st and 3rd bands both oscillate out of phase with ground(0th) while the 2nd band population always remains less then 0.1.
We also plot the autocorrelation function 
\begin{equation}
A(t)=|\langle \Psi(x,t=0)| \Psi(x,t) \rangle |^2
\label{eq:auto}
\end{equation}
in Fig.~\ref{fig:time-evolution}(d).
The peaks correspond to those of the ground band population in (c). The figures support the scenario of the time-independent picture above.

\begin{figure}[htbp]
 \begin{center}
 \includegraphics[width=7cm]{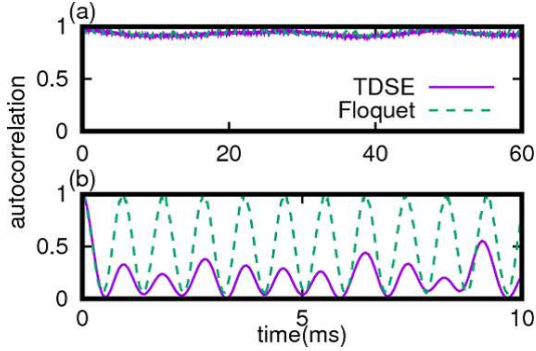}
 \end{center}
 \caption{(Color online) Autocorrelation functions with $\epsilon_0=0.165$ for (a) $E_\omega =$9.40 and (b)11.30, respectively.
 Green solid lines and purple dashed lines show results of the full TDSE calculations and those of Floquet theory respectively, the latter using a limited subset of Hilbert space.
 Floquet states discussed in subsection \ref{sect:floquet}. See text for details.
 }
 \label{fig:floquet}
\end{figure}

Let us corroborate the foregoing intuitive analysis and interpretation before going onto further exploration. 
We plot the autocorrelation function for two different excitation energies in Fig.~\ref{fig:floquet}.
In the case that the modulation frequency is far red-detuned, most of the wave packet remains localized in the ground band, thus the autocorrelation function $A(t)$ remains almost constant in time, Fig.~\ref{fig:floquet}(a).
When the modulation frequency exceeds the energy difference between ground and 2nd bands defined at $q=0$, then the Rabi cycle between 0th and 2nd band dominates, thus the autocorrelation function shows rapid oscillations, Fig.~\ref{fig:floquet}(b) with a markedly reduced amplitude for TDSE.
Importantly, a longer periodicity appears in $A(t)$ when the modulation frequency satisfies the closed cycle condition as seen in Fig.~\ref{fig:time-evolution}(d).

\begin{figure}[htbp]
 \begin{center}
 \includegraphics[width=7cm]{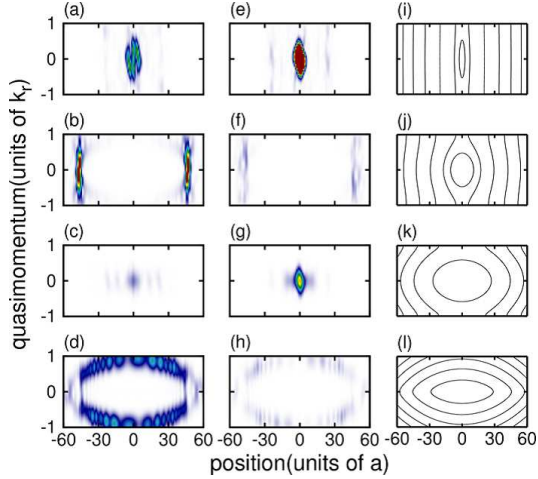}
 \end{center}
 \caption{(Color online) Snapshots of Husimi distribution functions $D(x,q,n,t)$ at two representative moments, (a)-(d) at t=7.9ms and (e)-(h) at t=15.7ms.
 Components corresponding to ground, 1st, 2nd and 3rd band from top to bottom are paired up side by side for the two moments.
 The group of figures (a)-(d) at t=7.9ms correspond to the first minimum of the autocorrelation function as in Fig.~\ref{fig:floquet}(b), therefore the most of the wave packet is placed in 1st band Bloch mode and 3rd band Dipole mode.
 In contrast, figures (e)-(h) at t=15.7ms correspond to the first maximum so that the wave packet is localized in the ground band with a noticeable amplitude in the second band.
 }
 \label{fig:husimi}
\end{figure}

Transforming the time-dependent wave function into the Husimi representation band by band, it is possible to construct quasi-classical phase-space probability distributions for each band.
This procedure gives a deep insight into the time-evolution of the current system from a classical mechanical viewpoint. 
Fig.~\ref{fig:husimi} thus shows Husimi distribution function $D(x,q,n,t)$ at t=7.9 and 15.7 ms with $E_\omega=10.2$.
The figures in small frames are arranged downward from the ground to the third band, and are paired side by side for the two moments t=7.9 and 15.7 ms.
We also plot the energy contours of the semi-classical Hamiltonian $H_{cl}$ in Fig.~\ref{fig:husimi}(i)-(l).
Fig.~\ref{fig:husimi} clearly shows that the wave function consists of ground-dipole, 1st-Bloch and 3rd-dipole components at the specified moments as we discussed. In this way, we surmise the sequence of transitions depicted in Fig.~ \ref{fig:time-independent}(b). We believe this particular wave packet ``{\it 4bandPWP}'' possesses such unique features that it merits a great deal of attention.

\subsection{Stability of ``{\it 4bandPWP}''}
\label{sect:stability}

\begin{figure}[htbp]
 \begin{center}
 \includegraphics[width=10cm]{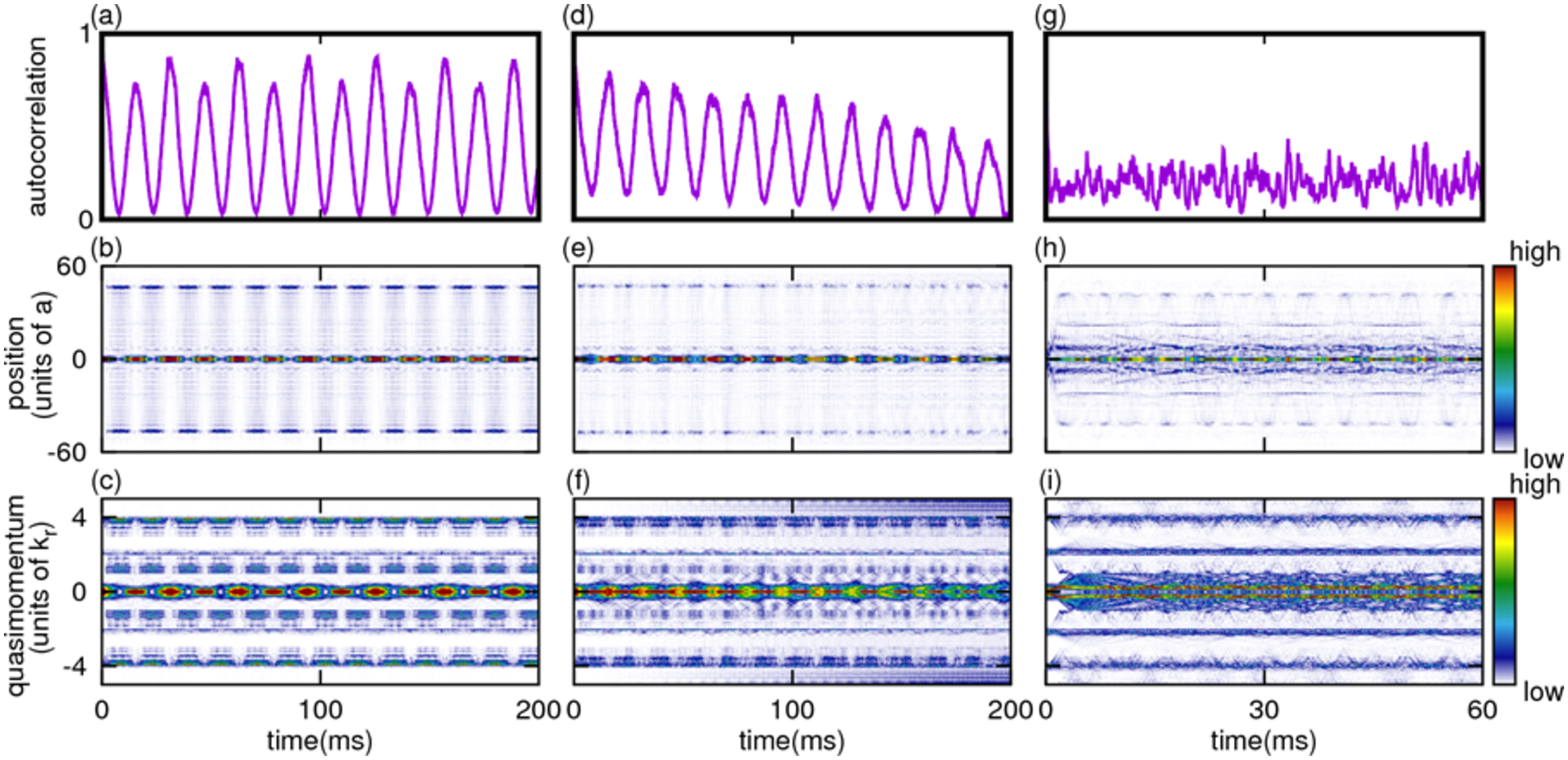}
 \end{center}
 \caption{(Color online)
  (a) Density plot of the autocorrelation function $A(t)$ defined by Eq.~(\ref{eq:auto}), (b)  in position and (c)  in quasimomentum space for $E_\omega=10.2$ and $\epsilon_0=0.165$ in a temporal range of $t$ up to 200 ms. 
  Likewise, (d), (e) and (f) are for $E_\omega=10.3$ and $\epsilon_0=0.165$ for the same temporal range. 
  Figures (g), (h), and (i) are shown for $t$ up to 60 ms for $E_\omega=10.2$, but for modulation amplitude raised to $\epsilon_0=0.4$. 
  The case $E_\omega=10.2$ is seen to be particularly stable while a slight variation of $E_\omega$ leads to rapid dephasing. 
  The stability region of {\it 4bandPWP} is restricted with respect to $E_\omega$ and $\epsilon_0$. 
 }
 \label{fig:compstab}
\end{figure}

We discuss stability of this {\it 4bandPWP} formed at $E_\omega=10.2$ first, and then go on to the Floquet analysis.
Various factors lead to loss of coherence in general. 
Among these factors are the finite lifetime of  eigenstates, the ensemble effect~\cite{ensemble}, disturbances from the environment and so on~\cite{suppress,ryd}.
Previously, inspired by experimental works~\cite{Hamburg}, we investigated the revival of Fermionic holes in a parabolic OL~\cite{my-2}. 
The holes, if perfectly isolated as assumed in our calculations thereof, would have infinitely long coherence  so that we were led to believe that the ensemble effect was the conceivable cause of dephasing in the actual experiment.
In this subsection, we study in much the same way the intrinsic stability of the present system under the perfect isolation condition.
Figs.~\ref{fig:compstab}(a)-(c) show results of the rigorous numerical simulation displayed in the same manner as in Fig.~\ref{fig:time-evolution}, but the time scale is extended.
The autocorrelation function shows temporally stable and periodic features in  (a), and also in the corresponding representations in (b) position and (c) quasimomentum space.
What makes the manifestation of {\it 4bandPWP} possible? Firstly, it is energetically confined below the 3rd band inclusive, thus involving only discrete-to-discrete transitions. 
Secondly, {\it 4bandPWP} is the only wave packet that satisfies the closed cycle condition starting from the neighborhood of $q=0$.
What happens if the modulation frequency is slightly detuned from $E_\omega$=10.2? We plot the results obtained at a slightly shifted modulation frequency, $E_\omega$=10.3 in Fig.~\ref{fig:compstab}(d)-(f) in the same manner as in (a)-(c).
In this case unlike that of $E_\omega=10.2$, the amplitude modulation brings the ground state up to and above the lowest tip of the 4th band, in other words, the closed condition is broken despite the smallness of the frequency shift.
The autocorrelation function shows departure from the seeming periodic motion.
It does not preclude the possibility of the ``quantum revival'' so that {\it 4bandPWP} may indeed revive over a much longer time span. We leave this topic for a separate occasion.
Fig.~\ref{fig:compstab}(g)-(i) shows the results with modulation amplitude raised to $\epsilon_0$=0.4.
Contrary to Fig.~\ref{fig:compstab}(a)-(c), the autocorrelation function is never close to unity.
The quasimomentum distribution in the ground band is strongly affected, and shows complicated structure.
The strong amplitude modulation violates the prerequisite that the amplitude modulation merely causes  vertical transitions in quasimomentum space.

\subsection{Analysis via Floquet eigenstates}
\label{sect:floquet}

The solution of TDSE can be expressed as a linear superposition of eigenstates $\{\chi_k (x)\}$ of Hamiltonian $H_0$, namely $\Psi(x,t)=\sum_k D_k(t) \chi_k (x)$.
Since the time-dependent term of $H$ is periodic in time, we apply the Floquet theory\cite{Floquet} to analyze and interpret our TDSE results.
In accordance with the Floquet theory\cite{Floquet-form}, the TDSE solution can be also represented as $\Psi(x,t)=\sum_\alpha C_\alpha \psi_\alpha (x,t)$, where $\{\psi_\alpha (x,t)\}$ is the set of Floquet eigenstates.
Each Floquet eigenstate can be decomposed into time-dependent and independent part so that factorizing out the dependence on  Floquet eigenenergy $E_\alpha$ leads to $\psi_\alpha (x,t)=e^{(-iE_\alpha t)}\displaystyle \sum_{m=-\infty}^{+\infty} e^{(-im E_\omega t)} F_m^\alpha (x)$ where $m$ is the number of  ``photons'' involved in the defining term.
Expanding the time-independent part $F_m^\alpha (x)$ in terms of the 0th order eigenstates $\{\chi_k (x)\}$ so that
 $F_m^\alpha (x) = \sum C_k^{m,\alpha} \chi_k (x)$, we obtain the following recurrence equation for coefficients $C_k^{m,\alpha}$,
\begin{equation}
(E_\alpha + m E_\omega ) C_k^{m,\alpha} = E_k C_k^{m,\alpha} + \frac{s\epsilon_0}{2} \sum_{k'} C_{k'}^{m+1,\alpha} \langle \chi_{k'} | \sin^2 (x)  | \chi_{k} \rangle + \frac{s\epsilon_0}{2} \sum_{k''} C_{k''}^{m-1,\alpha} \langle \chi_{k''} | \sin^2 (x) | \chi_{k} \rangle
\label{eq:reduced-F-ham}
\end{equation}
which is solved by diagonalization. 
To satisfy the initial condition,  we look for states which overlap significantly with the $m=0$ ground state among numerous eigenstates. 
The ``Floquet overlap coefficient''  $O_{\alpha}=|C_{k=0}^{m=0,\alpha}|$ for $k=0$ and $m=0$ is one measure of this overlap.
Now, to obtain sufficiently converged Floquet eigenenergies requires a large number of blocks labeled by the photon number $m$, possibly up to $m=30$ or so. For the purpose of qualitative discussions, however, we find $m=2$ adequate. In what follows, we thus restrict $|m|\le 2$ in diagonalization.  

\begin{figure}[htbp]
 \begin{center}
 \includegraphics[width=8cm]{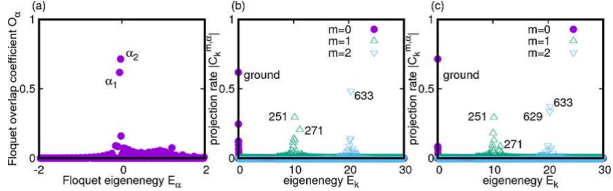}
 \end{center}
 \caption{(Color online)
  (a) Floquet overlap coefficient $O_{\alpha}=|C_{k=0}^{m=0,\alpha}|$ shows only two significant peaks labeled $\alpha_1$ and $\alpha_2$. 
  (b) Coefficient $|C_k^{m,\alpha}|$ as a function of $E_k$ for $\alpha_1$ and (c) for $\alpha_2$.
In (b), 0th order eigenstates of $k=251\ (m=1)$, $271\ (m=1)$ and $633\ (m=2)$ are markedly dominant. In (c), an additional component $k=629 \ (m=1)$ appears.  }
 \label{fig:projection}
\end{figure}

We show in Fig.\ref{fig:projection}(a) the distribution of the Floquet overlap coefficient for each eigenstate. 
The horizontal axis labels the eigenstates in the increasing order of their values. 
It is extremely noteworthy that we find {\it only two} eigenstates $\alpha_1$ and $\alpha_2$ with sizable Floquet overlap coefficient. 
Fig's \ref{fig:projection}(b) and (c) show their components identified by the eigenstates of $H_0$ and by the number of photons.
We may read off the time evolution of the two Floquet states from this information. 
Both are very similar, but they have some difference as we expand the details. The Floquet eigenstate in Fig.\ref{fig:projection}(b) consists mainly of the ground state, 251st(1st-band), 271st(bottom of 2nd band) and 633rd(3rd band) components while the one in Fig.\ref{fig:projection}(c) contains an additional component of  629th(3rd band).

Going back to Fig.~\ref{fig:floquet} for the plot of autocorrelation functions, the purple dashed lines represent those reconstructed with time-dependent wave functions of the two principal Floquet states.
In the resonant case (b), the result shows good qualitative agreement with the TDSE calculation, thus confirming our picture of the process depicted in Fig.\ref{fig:time-independent}, but the revival time deviates slightly because of the small cumulative energy shift due to
 non-resonant states. 

\subsection{The Four-Level model}
The periodic feature of the autocorrelation function is thus largely characterized by the energy difference between two principal Floquet states that satisfy the closed cycle condition.
We exploit a simple four-component model in order to analyze the period under the variation of such parameters as excitation frequency and modulation amplitude in an analytically tangible manner.  
The analysis will be guided by TDSE numerical results.

Now the four components consist of the ground,  second, third, and the fourth band. The two Floquet states are thus composed of these four under the following assumption.
One Floquet eigenstate consists of the ground and second band components, and the other one is a mixture of the first and the third.
The coupling strength $V_{kk^\prime}=s\epsilon_0 \langle \chi_{k'} | \sin^2 (x)  | \chi_{k} \rangle /2$ indeed introduces detuning to the resonant energy between ground and second band, thus providing a small energy shift to the ground state.
In addition, the mixed state of first and third band components being energetically resonant to the ground state,  it couples to the ground via the second band component.
Here we assume the quasi-eigenenergy of the Floquet state containing the ground band component is simply given in terms of the eigenenergies of $H_0$ by the following expression based on a $2\times 2$ sub-matrix, namely $E_A=\left[(E_0+E'_{k2})-\sqrt{(E_0+E'_{k2})^2 - 4 (E_0E'_{k2}-\epsilon_0^2 V_{0,k2}^2) } \right]/2$ where $E'_{k2}=E_{k2}-E_\omega$ corresponds to the dressed eigenenergy of the bottom of the second band.
The other quasi-eigenenergy is assumed likewise to be expressible as $E_B= \left[ (E'_{k1}+E''_{k3}) \pm \sqrt{(E'_{k1}+E''_{k3})^2 - 4 (E'_{k1}E''_{k3}-\epsilon_0^2 V_{k1,k3}^2) } \right]/2$ where $E'_{k1}=E_{k1}-E_\omega$ and $E''_{k2}=E_{k2}-2E_\omega$, respectively.
Note the presence of alternative branches here. Then the classical period is given by $T_r=\frac{2\pi}{|E_A-E_B|}$  
with $k1$, $k2$, and $k3$ suitably chosen, and with an appropriate choice of sign between the two terms in $E_B$.

\begin{figure}[htbp]
 \begin{center}
 \includegraphics[width=7cm]{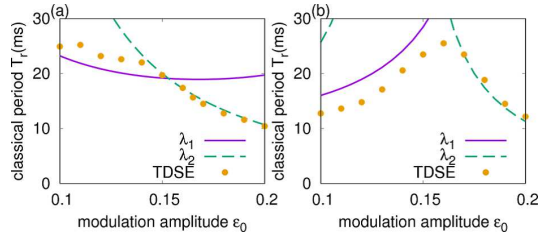}
 \end{center}
 \caption{(Color online) Classical period $T_r$ shown as a function of modulation amplitude $\epsilon_0$ for (a) $\nu=1.63\times 10^{-4}$ and (b) $\nu=3.32\times 10^{-4}$. Here $E_\omega$=10.2 for both (a) and (b).
 See text for the component values of the approximate Floquet eigenstates $\lambda_1$ (purple solid line) and $\lambda_2$ (green short dashed line).
 }
 \label{fig:revival}
\end{figure}

Fig.~\ref{fig:revival}(a) shows the classical period as a function of the modulation strength.
Here we choose two sets of eigenstates, $\lambda_1=(k1=251,k2=271,k3=633,-)$ and $\lambda_2=(251,271,629,+)$ where the number $k_i$ represents the $k_i$-th eigenstate of $H_0$ in our calculations, and the sign in the last column corresponds to that of $E_B$. 
These are an approximate representation of the two Floquet eigenstates that stand out in Fig.~\ref{fig:projection}.

One intriguing feature we wish to bring to the reader's attention is the smooth crossover from one Floquet state to another under the variation of the system's parameters $\epsilon_0$ and $E_\omega$.  
Fig.~\ref{fig:revival}(a) indicates that {\it 4bandPWP} indeed changes its character smoothly as a function of $\epsilon_0$ in accordance with the overlap of the two states $\lambda_{2}$ with respect to the state $\lambda_1$.
The dynamical feature of {\it 4bandPWP} appears to remain thus largely unchanged
 over this range of $\epsilon_0$.
The results of the model qualitatively agree with that of TDSE, however it deviates from $\lambda_1$ in lower side, possibly due to multi-photon type processes. 

In order to check the applicability of the model, we also applied the model to the case of $\nu=3.32 \times 10^{-4}$(100Hz) with $E_\omega=9.9$.
In (b), we plot the results in the same manner as in (a) with $\lambda_1=(170,191,425,+)$ and $\lambda_2=(167,191,425,+)$.
The suitable parameter set is chosen as discussed in the previous subsection.
The trend is almost the same as (a), showing the model works well in this regime.

\begin{figure}[htbp]
 \begin{center}
 \includegraphics[width=7cm]{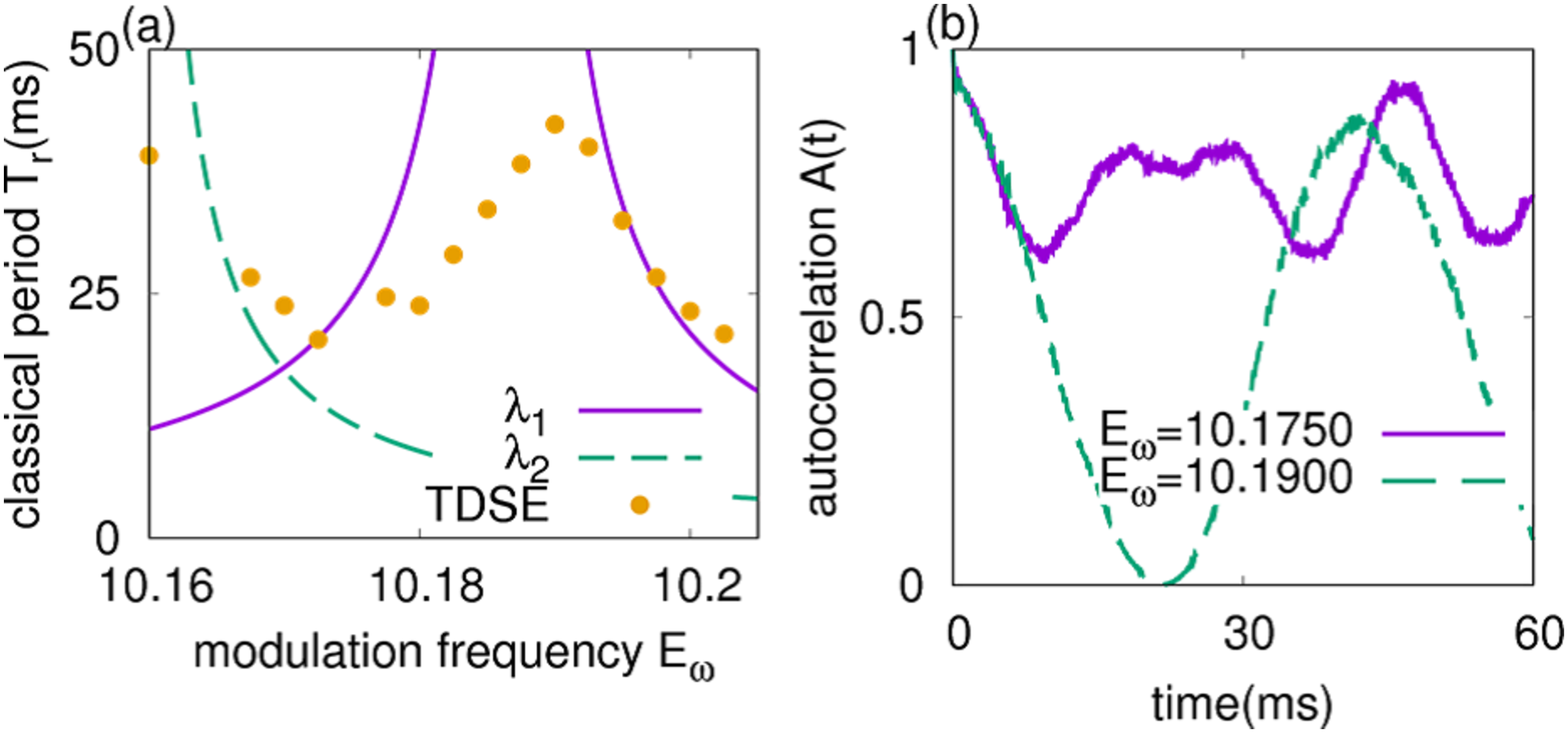}
 \end{center}
 \caption{(Color online) 
 (a) Dependence of classical period $T_r$ on modulation frequency $E_\omega$ with $\epsilon_0$ fixed to 0.12.
 The set of $\lambda_1$ and $\lambda_2$ in this case are the same as in Fig.~\ref{fig:revival}(a)
 The autocorrelation functions at specific modulation frequencies are shown in (b).
 Purple solid line and green short dashed line correspond to $E_\omega$=10.175 and 10.19, respectively.
 }
 \label{fig:revival_2}
\end{figure}

Next, we examine the modulation frequency dependence of {\it 4bandPWP} with $\nu=1.63 \times 10^{-4}$ and $\epsilon_0=0.12$.
Here in Fig.~\ref{fig:revival_2}, the suitable $\lambda$ set is such that $\lambda_1$ is the same as in Fig.~\ref{fig:revival}(a), but $\lambda_2=(251,271,629,-)$.
In the high frequency side $E_\omega\gtrsim$10.19, TDSE results follow primarily the curve of $\lambda_1$.
The depletion of ground state as shown in the autocorrelation $A(t)$, Fig.~\ref{fig:revival_2}(b), indicates a simple Rabi-oscillation between two states. On the other hand, the deviation of the TDSE result from this behavior at $E_\omega$ between 10.17 and 10.19 suggests presence of higher Rabi-oscillations due to the mixture of $\lambda_1$ and $\lambda_2$.

As we discussed above, the two comparable Floquet states appear to be reflected in the more complex oscillations of $A(t)$ at $E_\omega=10.1750$, namely at the intersection of $\lambda_1$ and $\lambda_2$.
Extending the model and analyzing this crossover phenomenon is of interest, but we leave this issue to future investigation. At any rate, the proposed model is thus seen to represent the main feature of {\it 4bandPWP} reasonably well.

\section{Conclusions}
\label{sect:conclusions}
In this paper, we extended our investigation of the dynamics of ultracold atomic wave packets in the amplitude modulated OL system\cite{Aarhus}. We sought rigorous numerical solutions of TDSE for the case of the continuously applied modulation. Among the obtained wave packets, we found a particularly intriguing one, the four-band periodic wave packet ``{\it 4bandPWP}''.  It jumps among four bands in such a way that its autocorrelation function exhibits stable periodicity, and covers a wide spatial range connecting the regions of localized and extended states. In bringing out these features, we considered the time-independent Hamiltonian, first with the aid of the phase-space map based on the energy-band structure, and then exploited a model based on the Floquet-theory using a limited Hilbert space. The model using a pair of principal Floquet solutions approximates {\it 4bandPWP} well, and reproduced the TDSE results very well.
This indicates that at or close to the closed cycle condition, only a few well-separated principal Floquet states dominate the dynamics of the system. 
In this regard, the Husimi representation proved also informative by providing temporal slices of the phase space of {\it 4bandPWP}, identifying moments when the inter-band transitions take place. 
A set of recommended parameter values is given in Table~\ref{tab:data} in hope of assisting future experimental investigations.  

\begin{table}[h]
\caption{A set of parameters used in the present simulation for realizing {\it 4bandPWP}.}
\begin{tabular}{ccccc}
\hline\hline
$s$      & $\nu$                     & $E_\omega$       & $\epsilon_0$ & $T_r$    \\
\hline
16$E_r$  &  $1.63\times 10^{-4}E_r$ & 10.2$E_r$       & 0.165        &  15.7 ms \\
\hline\hline
\end{tabular}
\label{tab:data}
\end{table}

Our results show that the periodic wave packet is sensitive to the modulation frequency. This point should be examined through future experiments.
Conversely, {\it 4bandPWP} may be used for calibrating the modulation frequency, or in other words,
this phenomenon can be pictured as a stable internal clock, and regarded as akin to well-controlled wave packet formation such as a pulse generation in the field of quantum optics.

\section*{Acknowledgments}
This work was supported by Research and Educational Consortium for Innovation of Advanced Integrated Science(CIAiS) and JSPS KAKENHI Grant Number 17K05596.


\begin{references}
\bibitem{ensemble}The ensemble effect here derives from nonuniform temperature variations of the system so that some classical trajectories lie within the separatrix while the rest lie outside. See \cite{my-2}.
\end{references}
\end{document}